# Dependence of balance energy on isospin degrees of freedom

S. Gautam[1], Aman D. Sood[1],* Rajeev K. Puri[1], Ch. Hartnack[2], and J. Aichelin[2]
[1]*Physics Department, Panjab University Chandigarh-160014, INDIA*
[2]*SUBATECH, Laboratoire de Physique Subatomique et des Technologies Associées,*
*Université de Nantes, IN2P3/CNRS-Ecole des Mines de Nantes,*
*4 rue Alfred Kastler, F-44072 Nantes, Cedex 03, FRANCE*
* email: amandsood@gmail.com

## Introduction

Collective transverse in-plane flow in heavy ion collisions has been a subject of intensive theoretical and experimental studies, as it can provide information about the nuclear matter equation of state (EOS) as well as in-medium nucleon-nucleon (nn) cross section [1]. The study of dependence of collective transverse flow on various entrance channel parameters as beam energy and impact parameter has revealed much interesting physics about the origin and properties of the collective flow. From these studies, it has been found that the transverse in-plane flow disappears at an incident energy termed as balance energy ($E_{bal}$) [2], where attractive part of the nuclear interactions balances the repulsive part. Presently, due to availability of the radioactive beams, role of isospin degrees of freedom in EOS can be studied. The collective transverse in-plane flow has been found to depend on isospin of the colliding system [3]. Here, we aim to study the dependence of $E_{bal}$ on N/Z ratio of the colliding system using IQMD model [4].

## Model

IQMD model is an extension of quantum molecular dynamics (QMD) model [1], which includes isospin degrees of freedom explicitly. The model consists of isospin-dependent Coulomb potential, symmetry potential, and nn cross sections. The total interaction potential is given by

$$V^{total} = \sum_{i}^{A}\frac{p_i^2}{2m_i} + \sum_{i}^{A}(V_i^{Skyrme} + V_i^{Yuk} + V_i^{Coul} + V_i^{mdi} + V_{sym}^{ij}), \quad (1)$$

where $V_i^{Skyrme}, V_i^{Yuk}, V_i^{Coul}, V_i^{mdi}$, and $V_{sym}^{ij}$ are respectively, the Skyrme, Yukawa, Coulomb, momentum dependent interaction (MDI), and symmetric potentials. The MDI is obtained by parameterizing the momentum dependence of real part of the optical potential. The final form of the potential reads as:

$$V^{MDI} = t_4 \ln^2(t_5(\vec{p}_1 - \vec{p}_2)^2 + 1)\delta(\vec{r}_1 - \vec{r}_2). \quad (2)$$

Here $t_4$ = 1.57 MeV and $t_5$ = 5×10$^{-4}$ MeV$^{-2}$. The symmetry potential between protons and neutrons corresponding to the Bethe-Weizsäcker mass formula has been included

$$V_{sym}^{ij} = t_6 \frac{1}{\rho_0} T_{3i} T_{3j} \delta(r_i - r_j), \quad (3)$$

where $t_6$ = 100 MeV, $T_{3i}$ and $T_{3j}$ denote the isospin projections of particles i and j.

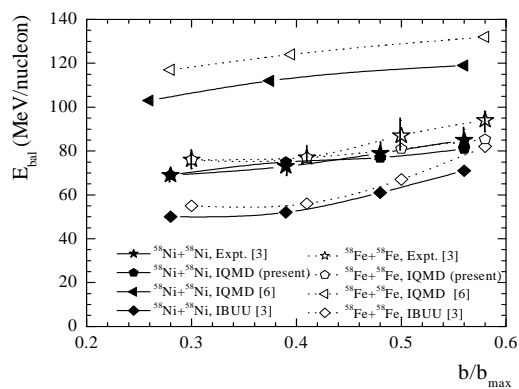

**Fig. 1** Balance energy as a function of reduced impact parameter. The solid (open) symbols are for $^{58}$Ni+$^{58}$Ni ($^{58}$Fe+$^{58}$Fe). The pentagons correspond to our preliminary results using IQMD model [4]. Stars represent the experimental data [3]. Diamonds and left triangles correspond to BUU [3] and IQMD [6] calculations, respectively. The values of $^{58}$Fe+$^{58}$Fe have been offset slightly in the horizontal direction for clarity. The lines are to guide eye.

### Results and discussion

We have simulated the reactions of $^{58}$Fe+$^{58}$Fe and $^{58}$Ni+$^{58}$Ni at incident energies ranging from 40 MeV/nucleon to 800 MeV/nucleon in small steps. The reactions were simulated at the reduced impact parameter bins of 0-0.28, 0.28-0.39, 0.39-0.48, and 0.48-0.56 (taken from Ref. [3]) using soft EOS with MDI (labeled as SMD) along with free nn cross section reduced by 20%. As shown in Ref. [5], the results with cross section reduced by 20-30% are in good agreement with the data. Fig. 1 shows $E_{bal}$ as a function of reduced impact parameter. Various symbols have been explained in the caption of fig. 1. Experimental values of $E_{bal}$ are plotted at upper limit of each impact parameter bin. The results of BUU model calculations [3] which takes care of isospin dependent potential as well as nn scattering cross sections along with IQMD calculations of Ref. [6] are also shown. From the figure, it is clear that our preliminary results matches much better with the data at all colliding geometries as compared to results of other theoretical model calculations [3, 6]. $E_{bal}$ for $^{58}$Fe+$^{58}$Fe has been found to be higher than that of $^{58}$Ni+$^{58}$Ni. This is because of the difference in cross sections for different nucleons [3, 6]. The neutron-proton cross section is approximately three times higher than proton-proton and neutron-neutron cross section. This leads to less repulsive collective flow in more neutron rich system, henceforth increasing the balance energy in neutron rich system.

### Acknowledgments

This work is supported by grant from Indo-French centre for the Promotion of Advanced Research under project no. IFC/4104-1/2009/1680.

### References

[1] H. Stöcker and W. Greiner, Phys. Rep. **137** (1986) 277; J.Aichelin, Phys. Rep. **202** (1991) 233; J. Achielin *et al*., Phys. Rev. Lett. **58** (1987) 1926.
[2] G. D. Westfall *et al*., Phys. Rev. Lett. **71** (1993) 1986; D. Krofcheck *et al*., Phys. Rev. Lett. **63** (1989) 2028.
[3] R. Pak *et al*., Phys. Rev. Lett. **78** (1997) 1022; Bao-An Li *et al*., Phys. Rev. Lett. **76** (1996) 4492
[4] Ch. Hartnack *et al.,* Eur. Phys. J. A **1** (1998) 151.
[5] D. J. Magestro *et al*., Phys. Rev. C **62** (2000) 041603.
[6] Chen Liewen *et al*., Phys. Rev. C **58** (1998) 2283.